\documentclass[final,1p,times,authoryear]{elsarticle}

\usepackage{amssymb}
\usepackage{amsmath}
 \usepackage{hyperref}
 \usepackage{color, soul}

\usepackage{rotating}

\journal{Earth and Planetary Science Letters}

\begin{document}

\begin{frontmatter}


\title{Metal Saturation and the Redistribution of Hydrogen in Earth’s Mantle} 

\author[inst1,inst2]{Junjie Dong\corref{cor1}}
\ead{dong2j@caltech.edu}
\cortext[cor1]{Corresponding author}
\author[inst3]{Lars P Stixrude}
\author[inst1]{Paul D Asimow}
\author[inst4]{Jie Li}

\affiliation[inst1]{organization={Division of Geological and Planetary Sciences, California Institute of Technology},
            city={Pasadena},
            postcode={CA 91125}, 
            country={USA}}
\affiliation[inst2]{organization={Department of Geosciences, State University of New York at Stony Brook},
            city={Stony Brook},
            postcode={NY 11794}, 
            country={USA}
            }
\affiliation[inst3]{organization={Department of Earth, Planetary, and Space Sciences, University of California},
            city={Los Angeles},
            postcode={CA 90095}, 
            country={USA}}
\affiliation[inst4]{organization={Department of Earth and Environmental Sciences, University of Michigan},
            city={Ann Arbor},
            postcode={MI 48109}, 
            country={USA}}

\begin{abstract}
Iron disproportionation reactions in mantle silicates can produce metallic iron that drives Earth’s deep mantle toward metal saturation under reduced conditions. Subducting slabs transport hydrated silicates to these depths, where interactions with metallic iron can reduce structurally bound hydrogen in silicates to reduced hydrogen-bearing phases, such as molecular hydrogen or iron hydrides, leaving mantle rocks in effect dry. Using the thermodynamic code HeFESTo with its latest self-consistent treatment of iron-bearing mantle phases, we investigate the stability and distribution of metallic iron in Earth's pyrolitic mantle across a broad range of oxidation states, represented by whole-rock Fe$^{3+}$/$\sum$Fe ratio from 1\% to 10\%. We find that metallic iron is present through much of the lower mantle across this range and, under very reduced compositions of whole-rock Fe$^{3+}$/$\sum$Fe = 1--3\%, extends into the upper mantle. Where subducted water meets metal-saturated regions, hydrous melts may form and migrate upward, rehydrating the overlying mantle or pooling near the transition zone. Metal saturation can thus redistribute hydrogen internally, creating a sharp contrast between a wet shallow mantle and a dry deep mantle. This redox-driven redistribution can decrease mantle silicate water storage capacity by 64--96\% today, to only 0.1--0.8 modern ocean masses, and may explain the viscosity contrast near the upper–lower mantle boundary. Although quantitative estimates of metal abundance and distribution depend on thermodynamic assumptions and remain uncertain above 50 GPa, our results reveal the role of redox reactions between disproportionated iron and subducted water in governing the speciation and redistribution of hydrogen in Earth's mantle.

\end{abstract}



\begin{keyword}{
    Iron disproportionation\sep	
    Metal saturation\sep		
    Mantle oxidation state\sep	
    Mantle hydrogen\sep
    Deep water cycle\sep
    Thermodynamic modeling
    }



\end{keyword}

\end{frontmatter}



\section{Introduction}
\label{intro}

Water on Earth's surface sustains its habitability over geological timescales. The abundance of surface water is regulated by fluxes in the deep water cycle between surface and interior, through volcanic outgassing and subduction recycling \citep[\textit{e.g.},][]{van_keken_subduction_2011}. In turn, the abundance of interior water alters the vigor of mantle convection and thereby influences these fluxes, as variations in mantle water content regulate both the rate of volcanic outgassing and the efficiency of water retention during subduction \citep[\textit{e.g.},][]{chotalia_coupled_2020}. 

While the abundance of surface water can be measured precisely, the abundance of mantle water, an interior reservoir that directly interacts with the surface, cannot. Geochemical constraints come from surface rock samples such as basalts and xenoliths, but they reflect only shallow mantle source regions. Remote sensing through electrical conductivity and seismic properties provides geophysical constraints on in situ mantle water content, but interpretations remain ambiguous: temperature, composition, and deformation can all mimic the effect of water, and resolution decreases with depth \citep[\textit{e.g.},][]{khan_earths_2016}. 

Laboratory measurements and thermodynamic modeling of structurally bound hydrogen (OH) solubility in the nominally anhydrous minerals (NAMs) within the ambient mantle can provide a theoretical upper bound for the bulk mantle water content, but reliable data are largely limited to upper mantle and transition zone minerals \citep[\textit{e.g.},][]{dong_constraining_2021}. For the lower mantle, available experimental data are scarce and inconsistent, often limited to the topmost lower mantle conditions (25--27 GPa), and scatter widely \citep[\textit{e.g.},][]{panero_dry_2015,fu_water_2019,liu_bridgmanite_2021,lu_substantial_2025}, ranging from nearly zero to a few thousand ppm by weight in bridgmanite, its most abundant mineral. As a result, both the form and abundance of water in the lower mantle, which makes up roughly half of Earth's mass, remain largely unresolved. 

Most existing experimental constraints on water-hosting mantle phases, from NAMs and hydrous minerals to hydrous melts, have been obtained in compositions more oxidized than those expected for Earth’s lower mantle. Compared with the shallow upper mantle, which has oxygen fugacity 
(\(f_{\mathrm{O}_{2}}\)) near the fayalite–magnetite–quartz (FMQ) buffer, the deep mantle is far more reduced, extending below the iron–wüstite (IW) buffer \citep[after][]{frost_experimental_2004,mccammon_paradox_2005,frost_redox_2008}. This reduced condition is imposed by the precipitation of metallic iron (Fe\(^{0}\)), produced by charge disproportionation of ferrous iron in mantle silicates. Such disproportionation reactions have been experimentally characterized in bridgmanite across lower mantle compositions \citep[][]{frost_experimental_2004,huang_composition_2021,huang_effect_2021,wang_decrease_2023,wang_bridgmanites_2025}, and in clinopyroxene and garnet in the deep upper mantle and transition zone under reduced compositions \citep[][]{rohrbach_metal_2007,beyer_reversed_2021}.

Once formed, disproportionated metal reacts with water-hosting phases typical of the oxidized shallow mantle, removing oxygen-bound hydrogen from NAMs, hydrous minerals, and melts, and forming reduced species such as molten iron hydride (FeH$_{x}$) and molecular hydrogen (H$_{2}$) fluid \citep[][]{yagi_iron_1995,okuchi_hydrogen_1997,shibazaki_hydrogen_2009,iizuka-oku_hydrogenation_2017,yuan_chemical_2018,zhu_metallic_2019,iizuka-oku_behavior_2021,kim_hydrogen-enriched_2023,zhu_deep_2025,zhu_iron_2025}. These iron--water reactions imply a fundamental dichotomy in mantle water distribution: in the metal-free shallow mantle, water is stored primarily as structural OH within the NAMs, together with hydrous minerals and melts carried by cold slabs; in contrast, the metal-saturated deep mantle disproportionates water, creating dry, oxidized rocks and modally minor, reduced hydrogen-bearing phases such as H$_{2}$ and molten FeH$_{x}$. These reduced hydrogen-bearing phases may migrate upward, rehydrating the overlying mantle and leaving the mantle rocks in the metal-saturated regions effectively dry. The transition from hydrated to dry silicates is ultimately controlled by the depth at which metal saturation is attained.

Despite the widespread occurrence of metal in the deep mantle, its role in the deep water cycle has not been systematically considered. Here, we investigate the distribution and abundance of metallic Fe\(^{0}\) in Earth's mantle and evaluate its role in governing the speciation and distribution of water across the full range of plausible mantle potential temperatures and oxidation states. We then discuss the geophysical implications, suggesting that hydrous melts responsible for seismic low-velocity zones near the transition zone may result from reactions between subducted water and the metal-saturated deep mantle, and that the viscosity contrast between the upper and lower mantle reflects hydrolytic weakening in the wet shallow mantle and its absence in the dry deep mantle. Our results are based on thermodynamic modeling and are therefore subject to uncertainties in experimental constraints on the ferric content of mantle minerals and in the choice of equations of state, particularly at lower mantle pressures; accordingly, we emphasize the robust qualitative trend in the prevalence of metallic iron in much of Earth’s lower mantle over geologic time, and discuss how its role in hydrogen redistribution can be further tested by experiments.

\section{Methods}
\label{methods}

\subsection{Iron disproportionation in ferric iron-bearing minerals.} 
We used the thermodynamic code HeFESTo \citep{stixrude_thermodynamics_2005,stixrude_thermodynamics_2011,stixrude_helmholtz_2024,stixrude_thermodynamics_2024} to compute the stable mineral assemblages of a pyrolitic mantle, including the abundance and distribution of metallic Fe$^{0}$ produced by disproportionation reactions. All calculations were based on the latest self-consistent parameter set, which incorporates the physics of iron-bearing mantle-related species \citep[]{stixrude_helmholtz_2024,stixrude_thermodynamics_2024}. 

Six ferric Fe$^{3+}$ species were considered across five mantle silicate minerals that involving charge disproportionation of iron:

\begin{enumerate}
    \item magnetite (Fe$^{2+}$Fe$^{3+}$Fe$^{3+}$O$_{4}$) in spinel (\textit{sp}), 
    \item acmite (Na$^{+}$Fe$^{3+}$Si$^{4+}$$_{2}$O$_{6}$) in clinopyroxene (\textit{cpx}),
    \item andradite (Ca$^{2+}$$_{3}$Fe$^{3+}$Fe$^{3+}$Si$^{4+}$$_{3}$O$_{12}$) in garnet (\textit{gt}),
    \item the Fe$^{3+}$Fe$^{3+}$O$_{3}$ and Fe$^{3+}$Al$^{3+}$O$_{3}$ components in bridgmanite (\textit{bg}) (Figure S1),
    \item and the Fe$^{3+}$Fe$^{3+}$O$_{3}$ component in post-perovskite (\textit{ppv}).
\end{enumerate} In previous versions of the HeFESTo code \citep{stixrude_thermodynamics_2005,stixrude_thermodynamics_2011}, only ferrous Fe$^{2+}$ was considered, and oxygen content was not explicitly defined but inferred from the number of cations. This fixed amount of oxygen, $b_{\mathrm{O}}^{*}$, was calculated as: 

\begin{equation}
    b_{\mathrm{O}}^{*} = 2b_{\mathrm{Si}} + b_{\mathrm{Mg}} + b_{\mathrm{Fe}} + b_{\mathrm{Ca}} + 1.5b_{\mathrm{Al}} + 0.5b_{\mathrm{Na}} + 1.5b_{\mathrm{Cr}},
\end{equation}
where all iron was assumed to be ferrous (Fe$^{2+}$). When multiple valence states of iron are introduced in \citet{stixrude_helmholtz_2024,stixrude_thermodynamics_2024}, the total amount of oxygen, $b_{\mathrm{O}}$, becomes an independent variable and \textit{no longer} equals $b_{\mathrm{O}}^{*}$. It is instead determined by the whole-rock Fe$^{3+}$/$\sum$Fe ratio, defined as: 
\begin{equation}
    2(b_{\mathrm{O}} - b_{\mathrm{O}}^{*})/b_{\mathrm{Fe}},
\end{equation}
where $b_{\mathrm{O}}^\ast$ represents the amount of oxygen assuming all iron is ferrous, and $b_{\mathrm{Fe}}$ is the total iron content. Full details on the treatment of iron disproportionation and several benchmark cases against experimental data are provided in the Supplementary Material, Sections S1--S4 and Figures S1--S8.\\

We assume a well-mixed mantle with a pyrolitic model composition and vary its oxidation state by adjusting the whole-rock Fe$^{3+}$/$\sum$Fe ratio \citep[][Table S1 in Supplementary Material]{workman_major_2005}. The bulk composition of pyrolite was taken from \cite{workman_major_2005} and includes SiO$_2$, MgO, FeO (total iron), CaO, Al$_2$O$_3$, Na$_2$O, and Cr$_2$O$_3$ (Table S1 in Supplementary Material). Estimates for various mantle rocks range from 2\% (xenoliths) to 6\% (MORB sources) \citep[\textit{e.g.,}][and references therein]{van_der_hilst_mantle_2005,frost_redox_2008,hirschmann_deep_2023,aulbach_mantle_2025}. We adopt a broader range of 1--10\% for the whole-rock Fe$^{3+}$/$\sum$Fe ratio in the pyrolitic mantle, and 1--3\% as the most likely present-day range \citep[\textit{e.g.,}][]{frost_experimental_2004,huang_composition_2021,wang_bridgmanites_2025}. To approximate the geothermal gradients of a convecting mantle, we calculated isentropes of mantle assemblages for potential temperatures ($T_\mathrm{p}$) ranging from 1600 to 1900 K. This range covers the thermal states of the solid mantle from the present-day to the early Archean \citep[\textit{e.g.},][]{herzberg_thermal_2010}.

\subsection{Water solubility in nominally anhydrous minerals.} 

When the ambient mantle is metal-free, the nominally anhydrous minerals (NAMs) become available to host structurally bound OH (and the maximum as water solubility). The bulk water storage capacity of mantle rocks, \(c_{\mathrm{H_2O}}^{\mathrm{mantle}}\), can be approximated, as an upper bound, by a weighted average (by their modal proportions \(X\)) of the water solubilities of the NAM phases present, \(c_{\mathrm{H_2O}}^{\mathrm{NAM}}\) at each pressure (\(P\)) and temperature (\(T\)) \citep[\textit{e.g.},][]{dong_water_2022}:

\begin{equation}
c_{\mathrm{H_2O}}^{\mathrm{mantle}}
= \left(c_{\mathrm{H_2O}}^{\mathrm{NAM}}\right)_{i}
\left(
X_{i} + \sum_{j} X_{j}\, D_{\mathrm{H_2O}}^{\,i/j}
\right)
\label{eq:bulk_capacity}
\end{equation}

\noindent \(D_{\mathrm{H_2O}}^{\,i/j}\) is the partition coefficient of H\(_2\)O between reference NAM phases \(i\), including olivine, wadsleyite, ringwoodite, bridgmanite, and post-perovskite) and other modally minor phases \(j\) including orthopyroxene, clinopyroxene, high-pressure clinoenstatite, garnet, akimotoite, stishovite, davemaoite, and ferropericlase.

For minerals with well-characterized pressure- and temperature-dependent solubility (olivine, wadsleyite, ringwoodite), we applied parameterized fits to the experimental data of the following equation \citep{dong_constraining_2021}:

\textbf{\begin{equation}
\ln\!\big(c_{\mathrm{H_2O}}\big)
= a
+ \frac{n}{2}\,\ln f_{\mathrm{H_2O}}(P,T)
+ \frac{\,b + c\,P\,}{T}
\label{eq:solubility}
\end{equation}}

\noindent Here, \(f_{\mathrm{H_2O}}(P,T)\) is the water fugacity (GPa), and the constants \(a\), \(b\), and \(c\) parameterize the entropy, enthalpy, and volume terms of the hydration reaction, respectively; \(n\) is the fugacity exponent. For the other phases, we used published partitioning coefficent of water relative to a reference NAM. Although the pressure–temperature dependence of water solubility in bridgmanite remains uncertain, our model is consistent with state-of-the-art experimental constraints with solubilities ranging from tens to hundreds of ppm by weight under oxidized conditions \citep[e.g.,][]{fu_water_2019,liu_bridgmanite_2021,lu_substantial_2025}. For details on the parameterization of NAM water solubility and data sources, see \cite{dong_constraining_2021,dong_water_2022}.

\section{Results and Discussion}
\label{results}

\subsection{Effects of pressure and temperature on ferric content in mantle minerals under reduced conditions}
\label{experiments}

To begin, we performed benchmark calculations against the available experimental constraints. We focus primarily on the effects of \(P\) and \(T\) on ferric content in mantle minerals in the reduced endmember of a pyrolitic mantle with a whole-rock Fe\(^{3+}\)/\(\Sigma\)Fe = 1\%. Such a reduced composition causes the mantle \( f\mathrm{O}_2 \) to stay at \(\Delta\)IW \(\approx-1\) anywhere deeper than 300 km and allows for the formation of metal from clinopyroxene, garnet and brigdmanite. This reduced endmember of pyrolite allows us to evaluate the performance of our model against experimental data collected under similarly reduced conditions, often imposed by metallic Fe buffers in multi-anvil (MA) experiments \citep[e.g.,][]{frost_experimental_2004,rohrbach_metal_2007,stagno_stability_2011,huang_composition_2021}. 

\begin{figure*}[!p]
    \centering
\includegraphics[width=0.7\linewidth]{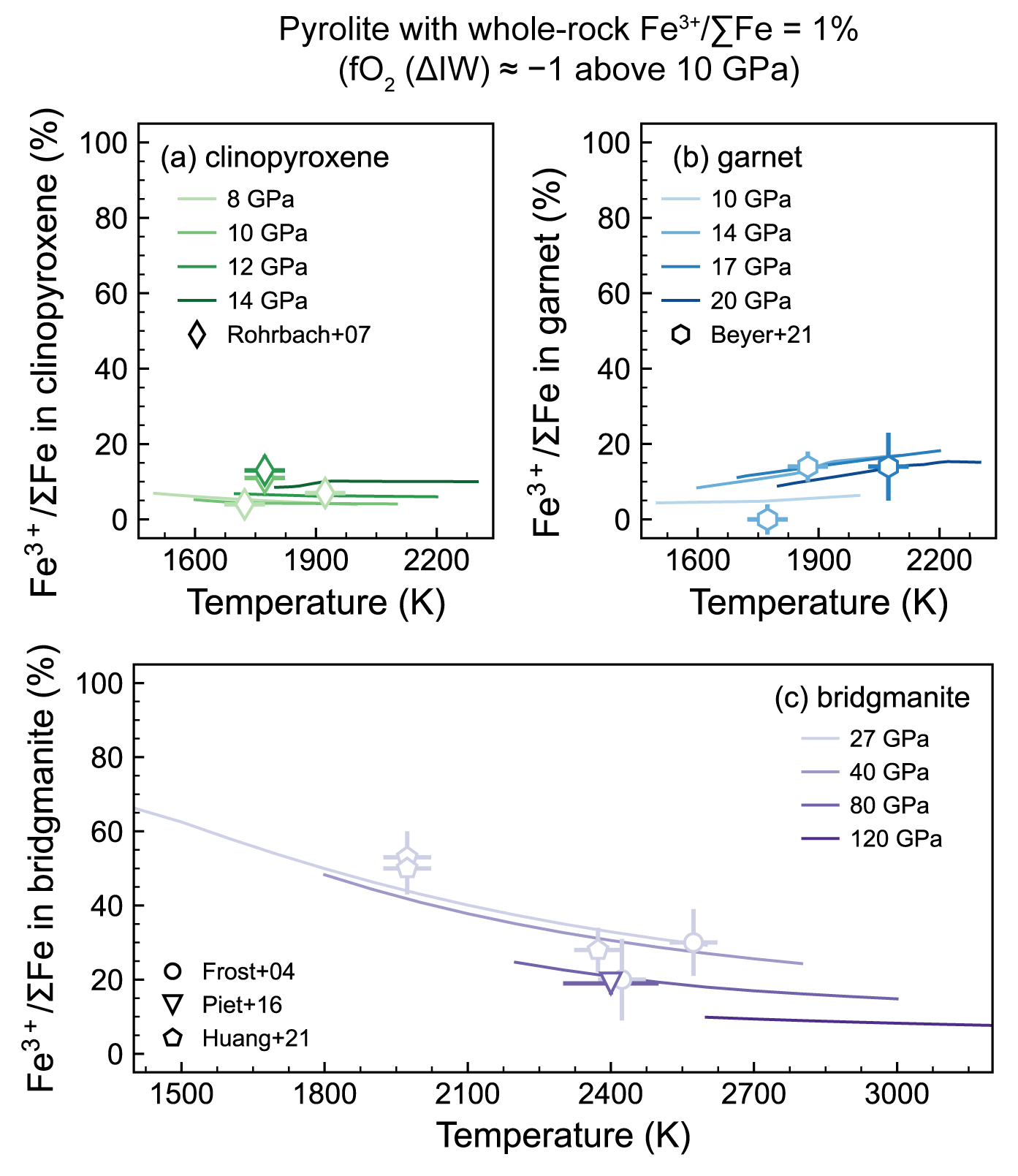}
    \caption{\textbf{\boldmath Predicted ferric content in mantle minerals within pyrolite at whole-rock Fe\(^{3+}\)/\(\sum\)Fe = 1\%, benchmarked against experimental data collected under similarly reduced conditions.} This composition represents a reduced endmember of Earth's mantle, in which metal saturation is present at all depths above 10 GPa with \(f\mathrm{O}_2 (\Delta\mathrm{IW}) \approx -1\). (\textbf{a}) Ferric content in clinopyroxene show negligible temperature dependence but increases slightly with pressure across 8--14 GPa; experimental data from \citet{rohrbach_metal_2007} (diamond) are shown for comparison. (\textbf{b}) Garnet shows a sutble positive temperature effect on its ferric content and a reversed pressure effect: ferric iron increases up to about 17 GPa before decreasing at 20 GPa, in alignment with the experimental data of \citet{beyer_reversed_2021}. (\textbf{c}) Ferric iron in brigmanite shows both negative temperature and pressure effects from 27 to 120 GPa, consistent with experimental data from \citet{frost_experimental_2004,huang_composition_2021}, and a reinterpreted data point from \citet{piet_spin_2016}. This reinterpreted data of \citet{piet_spin_2016} (at about 86 GPa and about 2400 K) is included as an upper bound for the reduced pyrolite (at whole-rock Fe\(^{3+}\)/\(\sum\)Fe = 1\% and \(f\mathrm{O}_2 (\Delta\mathrm{IW}) \approx -1\)). The negative pressure and temperature effects shown here align well with prior thermodynamic predictions of pressure dependence, which use independent choices of equation of state \citep{huang_effect_2021}, and recent experimental data to systematically constrain the temperature effect up to 2600 K \citep{wang_decrease_2023} and pressure effect up to 50 GPa \citep{wang_bridgmanites_2025}. See Section \ref{experiments} for details.}
    \label{fig:pyrolite_ferric_01}
\end{figure*}

Figures \ref{fig:pyrolite_ferric_01}a and \ref{fig:pyrolite_ferric_01}b show the \(P\)--\(T\) effects on the ferric content of clinopyroxene and garnet at \(f\mathrm{O}_2 (
\Delta\mathrm{IW}) \approx -1\) within pyrolite. No temperature effect was found for clinopyroxene, while a slight positive effect was predicted for garnet. By plotting different isobars, we also isolate and show the pressure effects for clinopyroxene and garnet. A positive pressure effect is predicted for clinopyroxene. Garnet initially shows a positive effect that is then reversed and followed by a negative effect between 17 and 20 GPa. These predictions broadly align with the experimental data of \citet{rohrbach_metal_2007, beyer_reversed_2021}, though the effects are subtle and require further experimental validation. Compared to the amount of metallic Fe\(^{0}\) that could be produced by bridgmanite in the lower mantle, the contribution of clinopyroxene and garnet in the upper mantle and transition zone is relatively small and limited to very reduced compositions with low whole-rock Fe\(^{3+}\)/\(\Sigma\)Fe (Figures. \ref{fig:fe_today}--\ref{fig:fe_3d}). Therefore, we will primarily focus on bridgmanite, and a comprehensive analysis of clinopyroxene and garnet within pyrolite under varying \(P\)--\(T\)--\(X\) conditions shall be pursued elsewhere.

Two ferric Fe$^{3+}$ species, Fe$^{3+}$Fe$^{3+}$O$_{3}$ and Fe$^{3+}$Al$^{3+}$O$_{3}$, are considered for iron disproportionation in bridgmanite. The latter is the dominant contributor to metal production (Figure S7). At \(f\mathrm{O}_2 (
\Delta\mathrm{IW}) \approx -1\), the ferric content of bridgmanite within pyrolite shows both negative temperature and negative pressure effects, agreeing well with the MA results at 25--27 GPa by \citet{frost_experimental_2004,huang_composition_2021} and aligned with our calculation of the Al-bearing bridgmanite with the simplified compositions (Figures. S3--S5). We highlight the negative pressure effect on bridgmanite that decreases metal production at greater depths using the isobars in Figure \ref{fig:pyrolite_ferric_01}c. This trend results from the leftward shift of the following reaction:

\begin{equation} \label{eq:fapv}
\underset{\mathrm{bridgmanite}}{\mathrm{Al}^{3+}\mathrm{Al}^{3+}\mathrm{O}_{3}} + 3\underset{\mathrm{ferropericlase}}{\mathrm{Fe}^{2+}\mathrm{O} } = \underset{\mathrm{metal}}{\mathrm{Fe}^{0}} + 2 \underset{\mathrm{bridgmanite}}{\mathrm{Fe}^{3+}\mathrm{Al}^{3+}\mathrm{O}_{3}}
\end{equation}

\noindent with increasing pressure and is ultimately governed by the equations of state (EoS) chosen for the species involved. This choice is particularly critical, as the FeAlO\(_{3}\) has not been synthesized as a pure perovskite-structured phase, and its EoS is inferred from measurements of bridgmanite solid solutions, the extrapolation of which may carry large uncertainties. We follow the EoS choices in \citet{stixrude_thermodynamics_2024}, supported by the results and analyses from \citet{dorfman_effect_2014}. This negative pressure effect for bridgmanite shown in Fig \ref{fig:pyrolite_ferric_01}c is consistent with previous thermodynamic models of Al-bearing simplified systems that used independent choices of equations of state for different bridgmanite species \citep{huang_effect_2021}, which reinforces the reliability of this negative pressure trend despite the uncertainties from different equation-of-state choices.

Benchmarking our calculations against experimental data, particularly the negative pressure effect on bridgmanite within a reduced pyrolitic mantle at \(f\mathrm{O}_2 (
\Delta\mathrm{IW}) \approx -1\) above 25--27 GPa, remains difficult, , primarily due to the scarcity of experimental constraints and their large uncertainties at lower mantle pressures. Almost all existing data relevant to lower mantle assemblages were collected under more oxidized conditions (\(f\mathrm{O}_2 (
\Delta\mathrm{IW}) > +1\)--\(+2\); \textit{e.g.}, \cite{piet_spin_2016,wang_bridgmanites_2025}) than what would be expected for a metal-saturated pyrolitic mantle with \(f\mathrm{O}_2 (
\Delta\mathrm{IW}) \approx -1\). 

\cite{wang_bridgmanites_2025} present the only set of MA experiments extending beyond 25--27 GPa and up to 50 GPa with well-characterized $f_{\mathrm{O}_{2}}$. Although these experiments are not metal-saturated and were conducted at $f_{\mathrm{O}_{2}}$ closer to \(\Delta\)IW \(\approx\) +2, rather than \(\Delta\)IW \(\approx\) -1, they can nonetheless be used to indirectly benchmark our thermodynamic model. Our calculations performed using simplified compositions, pressures, and temperatures identical to those of \cite{wang_bridgmanites_2025} show modest negative pressure effects on bridgmanite between 25 and 50 GPa, in excellent agreement with all experimental data reported in that study (Figures S4--S5). We emphasize that the pyrolitic composition adopted in Figure \ref{fig:pyrolite_ferric_01} is more complex than the simplified compositions often pursued experimentally \citep[\textit{e.g.},][]{wang_bridgmanites_2025}; pyrolite includes components such as CaO, Na$_2$O, and Cr$_2$O$_3$ and, for example, produces a realistic lower-mantle assemblage of bridgmanite, ferropericlase, and davemaoite, rather than simplified systems of (Fe, Al)-bearing bridgmanite with excess MgO or SiO$_2$ as is common in MA experiments \citep{frost_experimental_2004,huang_composition_2021,wang_decrease_2023,wang_bridgmanites_2025}. A full benchmark comparison is provided in the Supplementary Material, where we calculated the effects of \(P\)--\(T\)--\(X\) on bridgmanite using simplified compositions identical to those employed in the MA experiments.

In our calculations, the negative pressure effect becomes more pronounced above \(\sim\)80 GPa. In the absence of MA experiments above 50 GPa with well-characterized $f_{\mathrm{O}_{2}}$, we turn to diamond anvil cell (DAC) data for a semi-quantitative comparison. One DAC data point from Piet et al. (2016) shows a significantly lower ferric content in bridgmanite (\(19 \pm 3\) \% at \(2400\pm100\) K and \(86\pm4\) GPa), which is below both their other measurements and the whole-rock Fe\(^{3+}\)/\(\Sigma\)Fe of the starting material (\(25 \pm 4\) \%). Since bridgmanite is expected to host most of the ferric Fe\(^{3+}\) in lower mantle assemblages, its ferric content should exceed that of the bulk composition. This discrepancy suggests that the sample might have had more reduced conditions than originally reported, likely due to the open-system nature of DAC experiments, which do not use any redox buffer. We therefore include this reinterpreted data point in Figure \ref{fig:pyrolite_ferric_01}c as an upper bound for a reduced pyrolitic composition near 80 GPa. Under this reinterpretation, the negative temperature effect for bridgmanite at \(f\mathrm{O}_2 (
\Delta\mathrm{IW}) \approx -1\) that we predict can reasonably reproduce the data collected at \(86\pm4\) GPa) from \citet{piet_spin_2016}.

\subsection{Metal distribution in Earth’s deep mantle today}
\label{time}

\begin{figure*}
\centering
\includegraphics[width=1\textwidth]{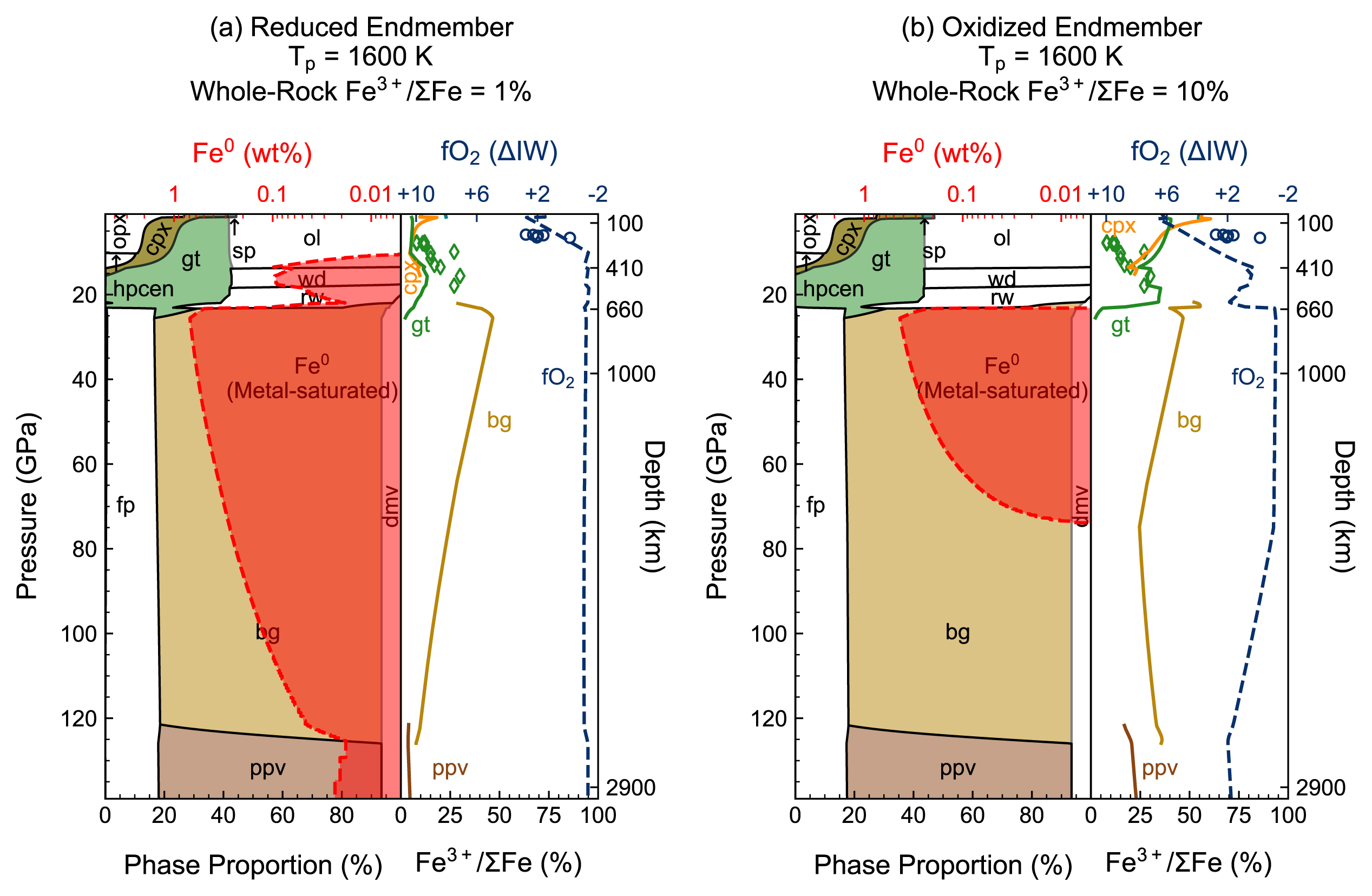}
\caption{\textbf{\boldmath Distribution of metallic Fe\(^0\) in the present-day mantle ($T_{\mathrm{p}}$ = 1600 K) for the reduced (1\%, a) and oxidized (10\%, b) endmembers.} (\textbf{a}) In the reduced endmember, metal saturation begins in the upper mantle, and Fe\(^0\) abundance decreases with depth as metal production from bridgmanite becomes less favorable at higher pressure. (\textbf{b}) In the oxidized endmember, metal saturation is confined to the upper- and mid-lower mantle and disappears above approximately 80 GPa as Fe is resorbed into mantle rocks. The contrast highlights how bulk mantle oxidation state (indicated by whole-rock Fe$^{3+}$/$\sum$Fe) controls both the depth and stability of Fe\(^0\) in the present-day Earth's mantle ($T_{\mathrm{p}}$ = 1600 K). Left panels show phase abundances as a function of depth, with mantle phases labeled (spinel = sp, orthopyroxene = opx, clinopyroxene = cpx, high-pressure clinopyroxeneolivine = hpcen, olivine = ol, wadsleyite = wd, ringwoodite = rw, bridgmanite = bg, ferropericlase = fp, davemaoite = dvm, post-perovskite = ppv). The lower $x$-axis gives cumulative phase proportions, with the red dashed line marking disproportionated metal; the upper $x$-axis shows the metal fraction on a logarithmic scale. Right panels show mantle oxygen fugacity $f$O$_{2}$ (navy dashed line) and ferric iron contents in individual minerals (colored lines). The lower $x$-axis gives the ferric iron content as a percentage of total iron in each phase, while the upper $x$-axis shows $f$O$_{2}$ normalized to the iron--wüstite (IW) buffer on a reversed logarithmic scale.}
\label{fig:fe_today}
\end{figure*}

With the thermodynamic model benchmarked against available experimental constraints, we now compute metal abundance in Earth's mantle across redox states and time. Here, we demonstrate how oxidation state influences metal saturation in the present-day mantle by modeling the most oxidized and most reduced endmembers of a pyrolitic mantle (whole-rock Fe$^{3+}$/$\Sigma$Fe = 1 and 10\%), which bracket the likely present-day range of 1--3\% \citep[\textit{e.g.,}][]{frost_experimental_2004,huang_composition_2021,wang_bridgmanites_2025}. For each endmember, we compute phase proportions of mantle mineral assemblages as a function of pressure and depth (Figure \ref{fig:fe_today}) along an isentrope with $T_{\mathrm{p}}$ = 1600 K.

For the reduced endmember (Figure \ref{fig:fe_today}a), metallic Fe$^{0}$ is stable from the upper mantle down to the core--mantle boundary. Its abundance increases to approximately 0.1 wt\% in the deep upper mantle and upper transition zone, where olivine and wadsleyite are predominant, and decreases to approximately 0.01--0.05 wt\% in the lower transition zone dominated by ringwoodite. This reversal is driven by the negative pressure effect on the disproportionation reaction in majoritic garnet at the transition zone pressures (Figure \ref{fig:pyrolite_ferric_01}b), where wadsleyite and ringwoodite (ferrous phases that coexist with garnet) favors resorption of metallic Fe$^{0}$ into the silicates, converting Fe$^{3+}$ and Fe$^{0}$ back to Fe$^{2+}$. This trend aligns with recent experimental data on garnet in KLB-1 peridotite and is consistent with the thermodynamic model of \citet[][]{beyer_reversed_2021}. At the top of the lower mantle, metallic Fe$^{0}$ increases again, peaking at approximately 0.69 wt\%, driven by disproportionation in bridgmanite. With increasing pressure, metal abundance decreases. Around 120 GPa, where bridgmanite transforms to post-perovskite, metallic Fe$^{0}$ falls to approximately 0.05 wt\%, with a similar amount subsequently produced by post-perovskite. 

In contrast, in the oxidized endmember in Figure \ref{fig:fe_today}b,  metallic Fe$^{0}$ is not stable in the upper mantle or transition zone. Iron disproportionation occurs only in bridgmanite and is confined to the lower mantle, where Fe$^{0}$ peaks at approximately 0.43 wt\% near its top and diminishes with depth. It is fully resorbed into bridgmanite at about 75 GPa and 1750 km. This stronger negative pressure effect is consistent with behavior predicted for more oxidized compositions (Figure S7) and is supported by the similar pressure dependence observed in simplified systems of bridgmanite with excess MgO (Figure S3--5). 

We compare these modeled endmembers with natural constraints, shown in Figure \ref{fig:fe_today}. The ferric Fe$^{3+}$ content of majoritic garnet in diamond inclusions \citep{kiseeva_oxidized_2018}, falls between our two endmember predictions. Likewise, the $f_{\mathrm{O}_{2}}$ values derived from Udachnaya mantle xenoliths \citep{miller_garnet_2016} align with our modeled range at pressures above 10 GPa, where the mantle geotherm becomes nearly adiabatic.

\subsection{Persistent metal saturation across redox states and time}

\begin{figure*}
\centering
\rotatebox{90}{%
\begin{minipage}{1\textheight} 
  \centering
  \includegraphics[width=1\textwidth]{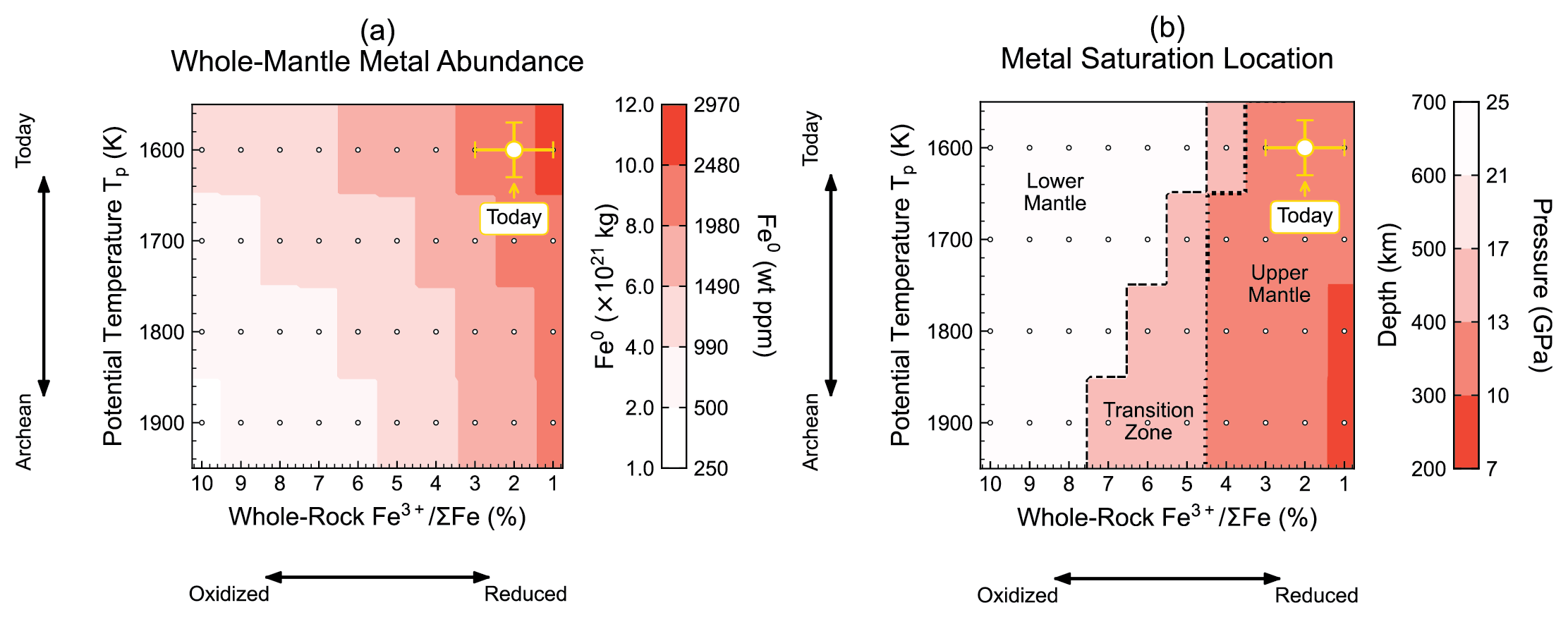}
  \caption{\textbf{\boldmath Metallic Fe$^{0}$ amounts to $8.6$–$10.5\times10^{21}$ kg (2120–2600 ppm by weight) in Earth’s mantle today, stabilizing near the base of the upper mantle and extending to the core–mantle boundary.} Whole-mantle metallic Fe$^{0}$ abundance (\textbf{a}) and saturation depth (\textbf{b}) are modeled as functions of potential temperature $T_{\mathrm{p}}$ (K) and oxidation state, expressed as the whole-rock Fe$^{3+}$/$\sum$Fe ratio (\%). Bulk metallic Fe$^{0}$ abundance (expressed in wt ppm and kg) increases with decreasing $T_{\mathrm{p}}$ and with increasing whole-rock Fe$^{3+}$/$\sum$Fe. For the present-day range of 1--3\%, metal saturation typically begins in the upper mantle (300--400 km). In more oxidized compositions (4--10\%), metal saturation is delayed to the transition zone and sometimes confined to the lower mantle. Higher $T_{\mathrm{p}}$ shifts the onset of metal saturation deeper, even as bulk metallic  Fe$^{0}$ abundance declines.}\label{fig:fe_all}
\end{minipage}}
\end{figure*}

To evaluate the prevalence of metal saturation and its systematic variation with mantle oxidation state over geological time, we extend our calculations across the full plausible range of mantle oxidation states, from whole-rock Fe$^{3+}$/$\sum$Fe = 1\% to 10\%, spanning the geological history of Earth's mantle since its solidification with mantle $T_{\mathrm{p}}$ = 1600 K (present day) to 1900 K (Archean). Our results show that metallic Fe$^{0}$ is widespread in the deep mantle across these ranges (Figure \ref{fig:fe_3d}) with its abundance and depth distribution depend on both whole-rock Fe$^{3+}$/$\sum$Fe and $T_{\mathrm{p}}$ (Figure \ref{fig:fe_all}). 

Mantle oxidation state, represented by the whole-rock Fe$^{3+}$/$\sum$Fe ratio, exerts the strongest control; more reduced mantles (i.e. lower whole-rock Fe$^{3+}$/$\sum$Fe) produce greater total metal mass (Table S2 in Supplementary Material; Figure \ref{fig:fe_all}a). At the present-day $T_{\mathrm{p}}$ of 1600 K and whole-rock Fe$^{3+}$/$\sum$Fe of 1--3\%, the total metal mass is about $8.6$–$10.5\times10^{21}$ kg (2120--2600 ppm by weight); under the same $T_{\mathrm{p}}$, but much more oxidized mantles (6--10\%, higher than present-day estimates), the total mass declines to about $4.0$–$5.1\times10^{21}$ kg (990–1260 ppm by weight). Oxidation state not only controls the total metal mass but also shift the depths at which metal saturation is attained. At constant $T_{\mathrm{p}}$, more reduced mantles reach metal saturation at shallower depths (Figure \ref{fig:fe_all}b). 
Under the assumption of a homogeneous mantle well mixed in whole-rock Fe$^{3+}$/$\sum$Fe, the combined negative dependence of pressure and temperature on bridgmanite lead to decreasing metal abundance with depth. The present-day mantle (1600 K, 1--3\%) is still likely to remain metal-saturated close to the core–mantle boundary. By contrast, more oxidized mantles than previously estimated develop a metal-free zone just above the core, leaving the mantle in disequilibrium with it. In reality, redox heterogeneity or stratification may drive the lowermost mantle closer to equilibrium with the core at the core–mantle boundary (CMB) \citep[][]{otsuka_deep_2012,henningsen_impactdriven_2025}, a boundary layer we have not modeled. 

Although less influential than mantle oxidation state, mantle $T_{\mathrm{p}}$ affects the depth and extent of metal saturation by simultaneously raising or decreasing pressure and temperature. Assuming constant whole-rock Fe$^{3+}$/$\sum$Fe, secular cooling of the mantle from $T_{\mathrm{p}}$ = 1900 K to 1600 K increases total metal mass from $7$ to $9.5\times10^{21}$ kg, raising the whole-mantle concentration from 1750 to 2340 ppm by weight. This increase is largely driven by the negative temperature dependence on iron disproportionation in bridgmanite (Figure \ref{fig:pyrolite_ferric_01}c; i.e. metal production from bridgmanite decreases with increasing temperature). Majoritic garnet, responsible for metal production in the upper mantle and transition zone, has a positive temperature dependence (Figure \ref{fig:pyrolite_ferric_01}b) opposite to that of bridgmanite (Figure \ref{fig:pyrolite_ferric_01}c) . Because bridgmanite dominates mantle mass, its negative temperature dependence tends to outweigh the positive temperature dependence of garnet, leading to a net decrease in whole-mantle metal abundance with higher $T_{\mathrm{p}}$.

\section{Implications}

\subsection{A dichotomy between wet shallow mantle and dry deep mantle}
\label{water}

Water reservoirs in Earth's mantle are conventionally understood to be dominated by structural water in NAMs and hydrous minerals, which can store substantial amounts of structurally bound OH under oxidized conditions near the FMQ buffer, for example, up to 2--3\% by weight in wadsleyite and ringwoodite in the transition zone \citep[\textit{e.g.},][]{dong_constraining_2021}. Bridgmanite, the most abundant mineral in Earth's interior and the dominant phase of the lower mantle, may also incorporate water, but reported measurements of its water solubility range widely, from nearly zero to several thousand ppm by weight \citep[\textit{e.g.},][]{panero_dry_2015,fu_water_2019,liu_bridgmanite_2021,lu_substantial_2025}. Assuming that all water is stored as structural OH in NAMs, including bridgmanite in the lower mantle, and that there are only negligible amounts of hydrous minerals due to their limited stability in the ambient mantle \citep[\textit{e.g.,}][]{yuan_stability_2019}, \cite{dong_constraining_2021} placed the maximum size of mantle water reservoirs at approximately 2.3 modern ocean masses (OM, where 1 OM = $1.335\times10^{21}$ kg) for the present-day mantle. This estimate is significantly reduced to 0.9 OM for a hotter early Archean mantle ($T_{\mathrm{p}}$ = 1900 K). 

\begin{figure*}
\centering
\begin{minipage}{1\textwidth} 
\centering
  \makebox[\textwidth][c]{%
\includegraphics[width=1\textwidth]{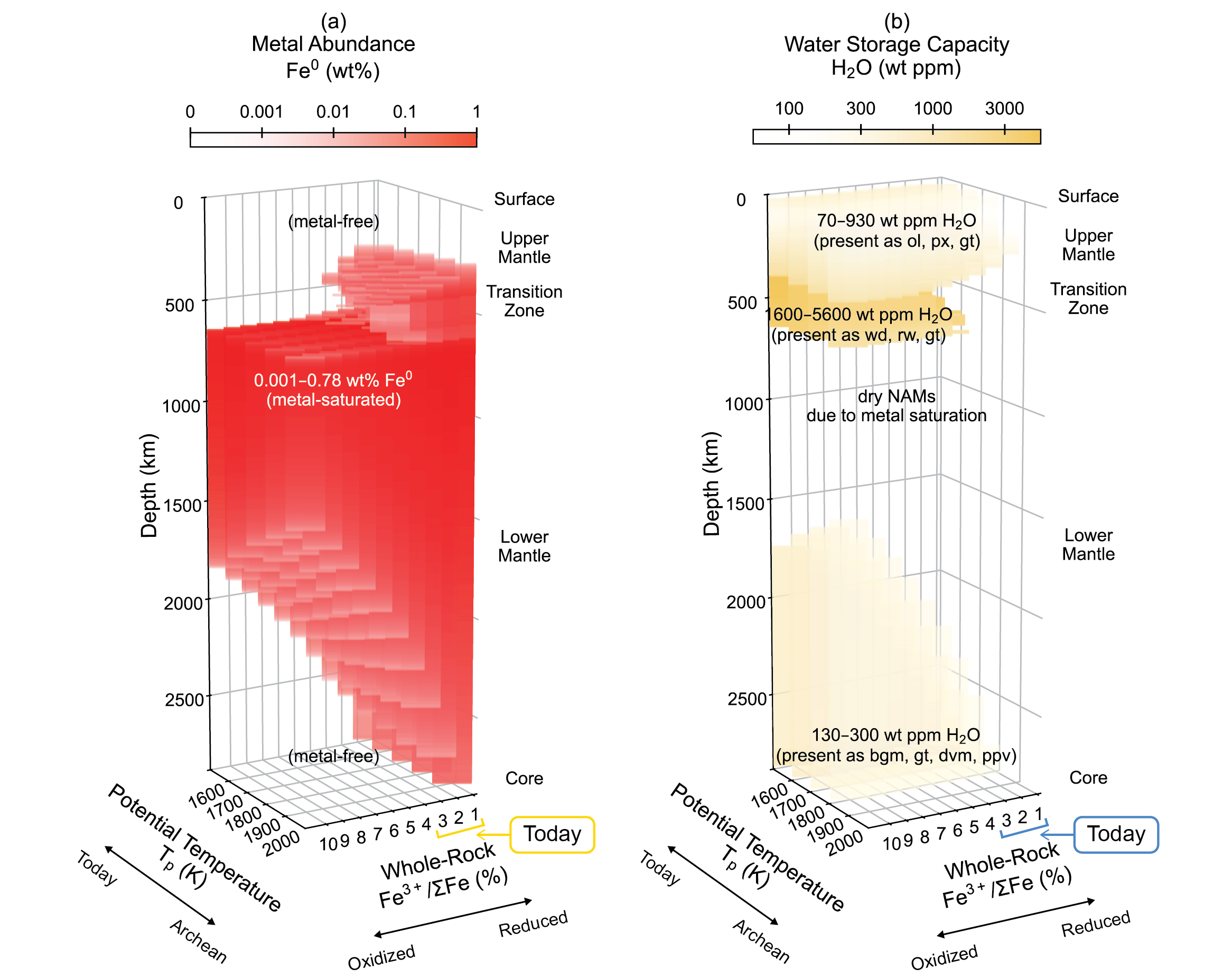}}
\caption{\textbf{Widespread metal saturation destabilizes structural OH in NAMs, resulting in a dry deep mantle.} Distribution of metal abundance (\textbf{a}) and water storage capacity (\textbf{b}) in Earth’s mantle are modeled as functions of depth (km), potential temperature $T_{\mathrm{p}}$ (K), and oxidation state, expressed as the whole-rock Fe$^{3+}$/$\sum$Fe ratio (\%). Present-day whole-rock Fe$^{3+}$/$\sum$Fe = 1--3\% yields persistent metal saturation from the upper to lowermost mantle, with Fe\(^0\) reaching up to 0.78 wt\% in (\textbf{a}). More oxidized bulk compositions (>3\%) confine metal saturation to the upper- and mid-lower mantle. NAMs in (\textbf{b}) can store up to 930 ppm H$_2$O by weight in the upper mantle, but transition-zone hydration is likely suppressed because metal saturation has been attained today with the whole-rock Fe$^{3+}$/$\sum$Fe range of 1--3\%. In contrast, in much more oxidized bulk compositions (6--10\%), NAMs can retain up to 5600 ppm H$_2$O; in the lowermost mantle (80--120 GPa), resorption of metal may once again permit a moderate H$_2$O storage capacity of 130--300 ppm.}
\label{fig:fe_3d}
\end{minipage}
\end{figure*}

However, such models presuppose a metal-free mantle and so exclude the effect of metal saturation formed from iron disproportionation, which becomes critical in Earth's deep mantle: unlike the relatively oxidized shallow mantle within log units of the FMQ buffer, the widespread occurrence of metallic Fe$^{0}$ imposes far more reduced conditions near the IW buffer and reacts with and consumes oxygen-bound hydrogen from NAMs, fundamentally redistributing water in the deep mantle. Experimental studies have consistently shown that, under mantle conditions, disproportionated metal can dehydrate water-hosting phases, including NAMs such as ringwoodite and wadsleyite \citep{shibazaki_hydrogen_2009,zhu_metallic_2019}, hydrous minerals such as brucite \citep{iizuka-oku_hydrogenation_2017,iizuka-oku_behavior_2021,kim_hydrogen-enriched_2023,zhu_deep_2025} and superhydrous phases B, E, and Egg \citep{zhu_deep_2025}, as well as hydrous melt and free water \citep{okuchi_hydrogen_1997,yuan_chemical_2018}, producing molten FeH\(_{x}\) \citep{yagi_iron_1995,iizuka-oku_behavior_2021,zhu_deep_2025,zhu_iron_2025} or molecular H\(_{2}\) fluid \citep{iizuka-oku_hydrogenation_2017,zhu_deep_2025}. These iron--water reactions take place at the margins of metal saturation in the deep mantle, which act as an ``iron wringer'' that extracts oxygen-bound hydrogen and prevents hydration of adjacent mantle rocks (Figure \ref{fig:fe_3d}b). Much of the deep mantle is then left dry in most modeled scenarios, despite having mineral assemblages with high intrinsic bulk water storage capacity under oxidized conditions.

As opposed to earlier models that considered only water solubility in NAMs \citep[\textit{e.g.},][]{dong_constraining_2021}, we explicitly account for the presence of disproportionated metal. Using the modeled metal abundance and distribution as a function of mantle $T_{\mathrm{p}}$ and whole-rock Fe$^{3+}$/$\sum$Fe (Figure \ref{fig:fe_3d}a), we evaluate the location of hydratable NAMs under oxidized mantle conditions and their bulk water storage capacity, excluding all reduced, metal-saturated zones (Figure \ref{fig:fe_3d}b; Section S5 in Supplementary Material). The results show that at present mantle conditions (1600 K, 1--3\%), relative to the metal-free model, the whole-mantle water capacity falls sharply by about 90\%, from 2.3 OM to 0.1--0.8 OM. For mantles far more oxidized than present-day estimates (6--10\%), this decrease is more modest at about 20\%. This notable contrast arises because, at low whole-rock Fe$^{3+}$/$\sum$Fe ratios, metal saturation extends up into the transition zone and even the upper mantle, preventing hydration in the region with the highest intrinsic water storage capacity when metal-free. At higher whole-rock Fe$^{3+}$/$\sum$Fe ratios, however, metal saturation remains largely confined to the lower mantle, thereby preserving significant bulk water storage capacity in the transition zone. Overall, depending on mantle $T_{\mathrm{p}}$ and whole-rock Fe$^{3+}$/$\sum$Fe ratio, the whole-mantle water storage capacity decreases from 0.7--2.3 to 0.1--1.9 OM, contracting by 18--96\% once metallic Fe\(^{0}\) from iron disproportionation is taken into account (Figure \ref{fig:water_all}; Table S3 in Supplementary Material). 

Figure \ref{fig:fe_3d}b shows how water storage capacity in NAMs varies with depth in relation to metal saturation. In the upper mantle at 10 GPa, NAMs can retain about 470 ppm water by weight, owing to the absence of metallic Fe\(^{0}\) at this depth. In the transition zone at 20 GPa, metal saturation eliminates water storage in NAMs under reduced bulk compositions (1--3\%), whereas under much more oxidized bulk compositions (6--10\%), NAMs retain a high water storage capacity (about 3900 ppm by weight). As shown in Figure \ref{fig:fe_3d}a, in the lowermost mantle between 80 and 120 GPa, our model predicts the loss of metal saturation, once again allowing moderate water storage of about 290 ppm by weight in bridgmanite. Our prediction is consistent with experimental observations: recent laser-heated diamond anvil cells experiments by \cite{zhang_pressure_2024} show that with 2\% water by weight, bridgmanite coexists with metallic Fe\(^{0}\) at 90 GPa but becomes metal-free above 105--108 GPa. 

Alternative choices of equations of state for bridgmanite, as discussed in Section \ref{experiments}, can allow metal saturation to extend all the way to the CMB at more oxidized bulk compositions than the whole-rock Fe$^{3+}$/$\sum$Fe ratio used here \citep{huang_effect_2021,wang_bridgmanites_2025}, eliminating any hydration of silicates at depth. The precise extent of metal saturation in the lowermost mantle remains to be tested experimentally; however, a weaker negative pressure effect does not negate, and may in fact reinforce, our central conclusion: for a wide range of plausible whole-rock Fe$^{3+}$/$\sum$Fe ratios, most of the deep mantle is predicted to have been metal-saturated throughout Earth’s history and, by extension, the metal-saturated lower mantle today likely contains only a negligible amount of structural OH. This reflects not the intrinsic water solubility in bridgmanite, if any under oxidized conditions \citep[\textit{e.g.},][]{fu_water_2019, liu_bridgmanite_2021,lu_substantial_2025}, but rather the incompatibility of its structural OH with disproportionated metal \citep[\textit{e.g.},][]{zhu_metallic_2019}.

\subsection{Redistribution of hydrogen in Earth's deep mantle and its geophysical consequences}
\label{discussion}

The final consideration is the consequences of iron–water reactions, which are expected to occur when subducting slabs deliver water-hosting phases into contact with metal-saturated deep mantle, and when ascending ambient mantle rises out of metal saturation into the oxidized shallow mantle. Experiments show that molecular H\(_2\) fluid and molten FeH\(_x\) are the two main hydrogen-bearing reaction products, with the dominant phase depending on whether water is in excess or metal is in excess (endmember reaction with excess water: \(2\mathrm{Fe} + 3\mathrm{H_2O} \rightleftharpoons \mathrm{Fe_2O_3} + 3\mathrm{H_2} \); endmember reaction with excess metal: \((1+\tfrac{2}{x})\mathrm{Fe} + \mathrm{H_2O} \rightleftharpoons \mathrm{FeO} + \tfrac{2}{x}\mathrm{FeH_x}\)) \citep{zhu_metallic_2019}. We infer that these same reactions also govern mantle-scale iron–water interactions. In Earth’s mantle, the likelihood of these two endmember reactions varies spatially and temporally. Regions dominated by sustained subduction and accumulation of hydrated slabs may locally reach water-excess conditions, favoring the production and rapid upward migration of H\(_2\) fluid. Conversely, FeH\(_x\) may form as the dominant reaction product when metal-saturated, deep-mantle assemblages ascend passively into shallower, more oxidized regions. FeH\(_x\) is unlikely to form an interconnected network in the silicate matrix and sink, given its extremely small modal abundance \citep{von_bargen_permeabilities_1986} and high interfacial tension with silicates \citep{hishinuma_surface_1994}, therefore it is expected to be trapped within reduced rock assemblages \citep{zhang_primordial_2016}. Upon gradual ascent into shallower depths, any FeH\(_x\) present in the rock assemblages would quickly destabilize as the ambient mantle \(f_{\mathrm{O}_{2}}\) rises above the IW buffer (Figure S10--S11 in Supplementary Material) and Fe and H\(_2\) become immiscible at shallow conditions \citep[e.g.,][]{stoutenburg_immiscibility_2026}.

The upward flux of either H\(_2\) or FeH\(_x\) has the potential to generate hydrous melts and rehydrate NAMs in the overlying mantle \citep[\textit{e.g.},][]{hirschmann_solubility_2012,yang_molecular_2016,yang_effect_2016,moine_molecular_2020,zhu_deep_2025}, assuming that H\(_2\) is oxidized before it can outgas to the atmosphere, as suggested by experimental constraints \citep[\textit{e.g.},][]{kohlstedt_diffusion_1998,mccammon_oxidation_2004,liu_solubility_2020}. Therefore, we propose that iron–water reactions primarily redistribute water internally within the mantle, resulting in water ultimately being stored in NAMs within the shallow, oxidized, metal-free region, leaving much of the deep mantle dry.  Iron-rich hydrous melts produced by iron--water reactions may pond at 410 km and 660 km depths \citep[\textit{e.g.},][]{huang_composition_2023}, depending on where metal saturation is encountered and neutral buoyancy is reached. These melts are consistent with seismic observations of low shear velocities at multiple locations and depths \citep[\textit{e.g.},][]{carr_high-resolution_2025}. Such features have been attributed to the ``water-filter'' model \citep{bercovici_whole-mantle_2003}, in which large contrasts in bulk water storage capacity across mantle silicates drive melt formation. Our model suggests an alternative: in the presence of metallic Fe\(^{0}\), silicate water solubility becomes effectively negligible, and hydrous melts can instead be generated by metal saturation as an ``iron wringer''. Both models seek to explain seismic low-velocity zones, but where the ``water-filter'' model emphasizes solubility contrasts among silicates, our ``iron wringer'' model emphasizes the reactive role of metallic iron.

Regardless of which reduced, hydrogen-bearing phase is dominant, our model reveals a sharp transition in the distribution of mantle water (or more precisely hydrogen) across the upper–lower mantle boundary: from a wet, shallow mantle above to a dry, deep mantle below. Such a non-uniform distribution of hydrated rocks could strongly influence the mantle viscosity profile. Observations indicate that the deep mantle viscosity is a factor of 30--100 higher than that in the shallow mantle \citep[after][]{peltier_new_1985}, with the transition likely occurring between the base of the upper mantle and the uppermost lower mantle. This dichotomy in water distribution offers a plausible explanation: in the shallow mantle, oxygen-bound hydrogen stored in NAMs significantly weakens silicates \citep[\textit{e.g.},][]{muir_water_2018}, whereas in the deep  mantle structural water is absent, with little hydrolytic weakening effect on silicates to first order. The widespread metal saturation in the deep mantle likely helps sustain this viscosity contrast by exerting strong redox control over where hydrated silicates can form, thus maintaining long-term non-uniform distribution in mantle hydrogen.

\section{Conclusions and Outlook}
\label{discussion}

Our thermodynamic modeling suggests that metallic iron, produced by iron disproportionation reactions, is present in much of Earth's deep mantle, from the Archean to the present, and across a broad range of oxidation states. Although the absolute abundance and exact depth distribution of this metal are sensitive to thermodynamic assumptions and experimental constraints, the predicted persistence of this exquisitely small amount of metal could govern the speciation and redistribution of hydrogen in the mantle. We therefore propose reframing the deep water cycle as a problem severely constrained by redox reactions rather than solely by water solubility in mantle minerals. We suggest that interactions between subducted water and disproportionated iron can redistribute hydrogen within the mantle internally, potentially producing sharp contrasts in water storage and, by extension, in rock rheology between the metal-free and metal-saturated regions. In this conceptual framework, hydrogen is preferentially retained in the shallow, more oxidized mantle, while the deeper mantle having liitle to no abilty to host structural OH in its silicates, even though the total amount of water or hydrogen in Earth's mantle remains unchanged in bulk.

Quantitative estimates at pressures above \(\sim\)50 GPa remain uncertain because the magnitude, and even the sign, of the combined pressure and temperature effects on metal saturation depends on the choice of equations of state for bridgmanite \citep{huang_effect_2021,stixrude_thermodynamics_2024,wang_bridgmanites_2025}. Improved constraints will require advances in redox-buffered experiments and precise chemical analyses under lower mantle conditions. While this uncertainty complicates quantitative estimates, it does not negate the persistence of metallic Fe$^{0}$ through much of the deep mantle over geological time, with profound implications for the coupling between mantle redox state and the deep water cycle. 

\bibliographystyle{elsarticle-harv} 
\bibliography{refs}



\clearpage

\clearpage

\clearpage

\clearpage

\clearpage
\begin{figure*}
\centering
\rotatebox{90}{%
\begin{minipage}{1\textheight} 
  \centering
  \includegraphics[width=1\textwidth]{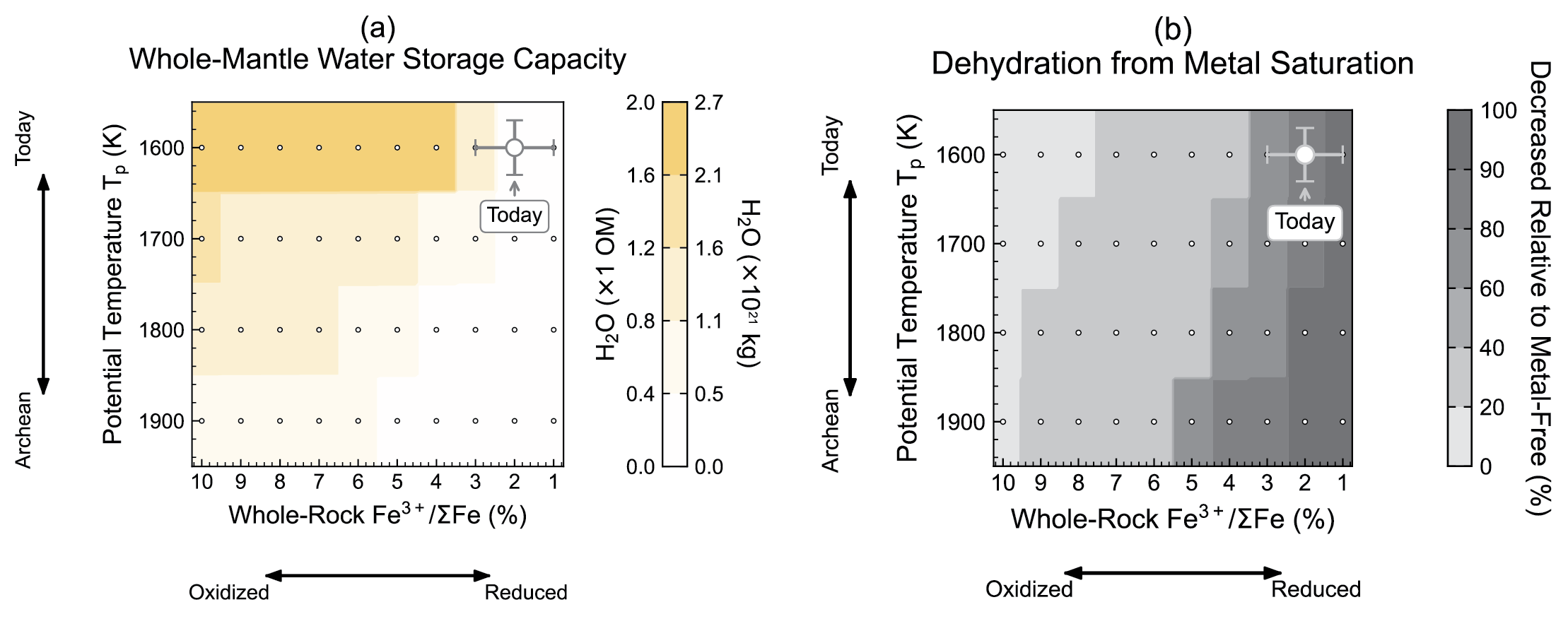}
  \caption{\textbf{Whole-mantle water storage capacity falls by 64--96\% today, from 2.3 to 0.1--0.8 modern ocean masses, when metal saturation is taken into account.} Whole-mantle water storage capacity with metal saturation (this work, \textbf{a}) are modeled as functions of potential temperature $T_{\mathrm{p}}$ (K) and oxidation state, expressed as the whole-rock Fe$^{3+}$/$\sum$Fe ratio (\%), and its decrease relative to metal-free models \citep[\textbf{b},][]{dong_constraining_2021}. Whole-mantle water storage capacity (expressed in kg and modern ocean mass, OM) increases with decreasing $T_{\mathrm{p}}$ and with increasing whole-rock Fe$^{3+}$/$\sum$Fe. For the present-day range of 1--3\%, whole-mantle storage capacity ($T_{\mathrm{p}}$ = 1600 K) is reduced by 64--96\% due to metal saturtion, from nearly $3.00\times10^{21}$ kg (2.3 OM, \cite{dong_constraining_2021}) to $0.12$--$1.08\times10^{21}$ kg (0.1--0.8 OM, this work). In more oxidized compositions (4--10\%), present-day storage capacity has a more modest decrease relative to metal-free models (18--27\%). The larger drops in storage capacity occur when lower whole-rock Fe$^{3+}$/$\sum$Fe and higher $T_{\mathrm{p}}$ enable metal saturation in the transition zone, where silicates would otherwise have the highest intrinsic water storage capacity under metal-free conditions.}\label{fig:water_all}\end{minipage}}
  \end{figure*}

\clearpage

\end{document}